\newif\ifOneCol

\ifOneCol
	\documentclass[draftcls,onecolumn,11pt]{IEEEtran}
\else
	\documentclass[twocolumn,10pt]{IEEEtran}

\fi
\usepackage{cite}
\usepackage[pdftex]{graphicx}

\pdfoutput=1
\usepackage[cmex10]{amsmath}
\usepackage{amsfonts,amssymb}
\usepackage{verbatim}
\usepackage{color}
\usepackage{multirow} 
\usepackage{makecell} 

  \graphicspath{{graphics/}}
  \DeclareGraphicsExtensions{.pdf,.jpeg,.png}

\usepackage{algorithm}
\usepackage{algpseudocode}

\usepackage{array}

\usepackage{mdwmath}
\usepackage{mdwtab}
\DeclareMathOperator{\erf}{erf}

\newcommand{\numDim}{\mathcal{D}}
\newcommand{\x}{x}
\newcommand{\y}{y}
\newcommand{\z}{z}
\newcommand{\xEff}{\hat{x}}
\newcommand{\yEff}{\hat{y}}
\newcommand{\zEff}{\hat{z}}
\newcommand{\zI}{\hat{x}_\mathrm{i}}
\newcommand{\zF}{\hat{x}_\mathrm{f}}

\newcommand{\indMolType}{m}

\newcommand{\Dx}[1]{D_{#1}}
\newcommand{\dist}{d}
\newcommand{\Nx}[1]{N_{#1}}
\newcommand{\Nxtavg}[2]{\overline{\Nx{}}_\textnormal{#2}\left(#1\right)}

\newcommand{\Cxtavg}[2]{\overline{C}_\textnormal{#2}\left(#1\right)}
\newcommand{\Nxt}[2]{{\Nx{}}_\textnormal{#2}\left(#1\right)}
\newcommand{\Cx}[1]{C_{#1}}

\newcommand{\posVec}[1]{\vec{l}_{#1}}
\newcommand{\posScalar}[1]{l_{#1}}
\newcommand{\flowVec}{\vec{v}}

\newcommand{\flowScalar}[1]{v_{#1}}
\newcommand{\flux}[1]{\mathbf{J}_{#1}}
\newcommand{\Peclet}{Pe}

\newcommand{\EXP}[1]{\exp\left(#1\right)}
\newcommand{\ERF}[1]{\erf\left(#1\right)}

\newcommand{\numTrials}{n}
\newcommand{\numSuccess}{k}

\newcommand{\trialProb}{p}
\newcommand{\varOne}{X}

\newcommand{\subLength}[1]{h_{#1}}

\newcommand{\normRV}[1]{n_{#1}}
\newcommand{\uniRV}[1]{u_{#1}}

\newcommand{\dtMicro}{\Delta t}

\newcommand{\timeX}[1]{t_{#1}}

\newcommand{\prop}[1]{\alpha_{#1}}

\newcommand{\dtInf}{\delta t}

\newcommand{\setSub}{\Omega_\textnormal{Sub}}

\newcommand{\numMolSubX}[1]{U_{#1}}
\newcommand{\areaOverlap}{A_\textnormal{o}}
\newcommand{\transRate}[1]{k_{#1}}

\newcommand{\meter}{\textnormal{m}}
\newcommand{\second}{\textnormal{s}}

\newcommand{\RX}{\textnormal{RX}}
\newcommand{\threeD}{\textnormal{3D}}
\newcommand{\oneD}{\textnormal{1D}}

\newcommand{\peak}{\mathrm{p}}

\newcounter{mytempeqncnt}

\newcommand{\edit}[2]{#1}

\begin{document}

%
\title{Algorithm for Mesoscopic Advection-Diffusion}

\author{Adam Noel, \IEEEmembership{Member, IEEE}, and Dimitrios Makrakis
	\thanks{Manuscript submitted 31 May, 2018. Revision submitted 11 September, 2018. This work was supported in part by the Natural Sciences and Engineering Research Council of Canada (NSERC) through a postdoctoral fellowship.}
	\thanks{A.~Noel is with the School of Engineering, University of Warwick, Coventry, UK, (email: adam.noel@warwick.ac.uk).}
	\thanks{D.~Makrakis is with the School of Electrical Engineering and Computer Science, University of Ottawa, Ottawa, ON, Canada (email: dimitris@eecs.uottawa.ca).}}

	\maketitle

\begin{abstract}
In this paper, an algorithm is presented to calculate the transition rates between adjacent mesoscopic subvolumes in the presence of flow and diffusion. These rates can be integrated in stochastic simulations of reaction-diffusion systems that follow a mesoscopic approach, i.e., that partition the environment into homogeneous subvolumes and apply the spatial stochastic simulation algorithm (spatial SSA). The rates are derived by integrating Fick's second law over a single subvolume in one dimension (1D), and are also shown to apply in three dimensions (3D). The proposed algorithm corrects the derived rates to ensure that they are physically meaningful and it is implemented in the AcCoRD simulator (Actor-based Communication via Reaction-Diffusion). Simulations using the proposed method are compared with a naive mesoscopic approach, microscopic simulations that track every molecule, and analytical results that are exact in 1D and an approximation in 3D. By choosing subvolumes that are sufficiently small, such that the P\'{e}clet number associated with a subvolume is sufficiently less than 2, the accuracy of the proposed method is comparable with the microscopic method, thus enabling the simulation of advection-reaction-diffusion systems with the spatial SSA.
\end{abstract}


\section{Introduction}

Stochastic simulations play an important role to help model and understand biochemical and biophysical systems at small physical scales. They can be used to support or predict both analytical models and experimental results. Prominent areas of application include biological cell signaling and the design of molecular communication systems; see \cite{Andrews2010,Farsad2016}, respectively.

Stochastic simulations of physical chemistry occupy a middle ground between rigorously modeling all individual molecules (i.e., a \emph{molecular dynamics} approach) and modeling all molecules with continuous concentrations. \edit{Molecular dynamics modeling is computationally expensive and can be unsuitable for systems above a nanometer scale. Modeling concentrations as continuous is computationally efficient but cannot capture micrometer scale behavior where local concentrations can deviate significantly from their expected values. The compromise made by stochastic simulations is to}{R2C2} represent the medium as a continuous solvent and model dilute solute molecules as discrete populations. The stochasticity applies to the corresponding behavior of the solute due to the sparsity of the individual molecules. For example, the displacement of a molecule via collisions with solvent molecules (i.e., diffusion) can be modeled as a Gaussian random variable (see \cite[Ch.~1]{Berg1993}), and the stochasticity of the chemical master equation has been shown to exactly represent the reaction dynamics of a well-stirred system in thermal equilibrium (see \cite{Gillespie1992}).

A common categorization of stochastic reaction-diffusion simulators is as either \emph{microscopic} or \emph{mesoscopic}, \edit{as summarized in Table~\ref{table_simulators}}{R2C3}. Microscopic simulators, such as Smoldyn \cite{Andrews2004} and BiNS2 \cite{Felicetti2013}, track each solute molecule individually. This approach is typically discrete in time and continuous in space, such that the system evolves according to a global time step. Mesoscopic simulators, such as MesoRD \cite{Elf2004} and URDME \cite{Drawert2012}, partition the environment into virtual containers and track the molecule population in each container (or \emph{subvolume}). This approach is discrete in space and continuous in time. Generally, the improved spatial resolution makes microscopic simulators more accurate but also more computationally intensive. The trade-offs in accuracy and computational complexity have motivated the development of hybrid simulators that separate the environment into microscopic and mesoscopic domains; see \cite{Flegg2014}. Examples of simulators that have implemented a hybrid approach include URDME \cite{Hellander2012}, Smoldyn \cite{Robinson2015}, and our platform AcCoRD \cite{Noel2017a}. \edit{Further discussion comparing these approaches can be found in \cite{Noel2017a}.}{R2C3}

Diffusion and chemical reactions are not the only relevant phenomena in biochemical and biophysical systems. Many fluid environments also have a bulk flow (or ``drift'') associated with them, e.g., blood flow and air currents. Advection biases diffusion in some direction and could vary locally or over time. Implementing flow in a microscopic model is relatively simple; every execution of Brownian motion includes a deterministic bias in the direction of the flow; see \cite[Ch.~4]{Berg1993}. However, in the mesoscopic approach, the implementation of flow is non-trivial. 

In this paper, we propose an algorithm to implement flow in mesoscopic reaction-diffusion simulations. The core of a mesoscopic simulation is the Stochastic Simulation Algorithm (SSA), which was originally proposed by Gillespie in \cite{Gillespie1977} and is also known as Gillespie's algorithm. The SSA generates exact trajectories of the evolution of a homogeneous chemical system. Diffusion is included by partitioning the physical space into homogeneous subvolumes, such that diffusion between subvolumes is treated as a special type of reaction. Every possible ``reaction'' (including diffusion) is assigned a corresponding ``reaction rate,'' and the algorithm is known as the spatial SSA. The primary contribution of this work is to properly modify the rates associated with diffusion reactions to account for the addition of a steady (but not necessarily uniform) drift.

To the best of our knowledge, advection has not been fully integrated within a mesoscopic reaction-diffusion simulator. The implementation of advection in \cite{Hellander2010} considers mesoscopic flow only in the absence of diffusion, which simplifies the modification of the transition rates. The advection-diffusion model in \cite{Ancey2015} leads to observations that are similar to what we can obtain using our approach, but it uses a Langevin method, which does not have the temporal precision of the spatial SSA.

\begin{table}[!tb]
	\centering
	\caption{\edit{Summary of stochastic reaction-diffusion approaches.}{}}
	
	{\renewcommand{\arraystretch}{1.2}\footnotesize
		\begin{tabular}{l||l|m{4.5cm}}
			\hline
			\bfseries Approach & \bfseries Examples & \bfseries Features  \\ \hline \hline
			Microscopic & \makecell{Smoldyn \cite{Andrews2004}\\ BiNS2 \cite{Felicetti2013}} & Track individual molecules; generally better spatio-temporal resolution. \\ \hline
			Mesoscopic & \makecell{MesoRD \cite{Elf2004} \\ URDME  \cite{Drawert2012}} & Track molecules in subvolumes; generally better computational efficiency; better scalability potential. \\ \hline
			Hybrid & \makecell{URDME \cite{Hellander2012} \\ Smoldyn \cite{Robinson2015} \\ AcCoRD \cite{Noel2017a}} & Simulations can have both microscopic and mesoscopic regions. \\ \hline
		\end{tabular}
	}
	\label{table_simulators}
\end{table}

Our goal is to broaden the applicability of mesoscopic simulations (or hybrid simulations with a mesoscopic component) by introducing the ability to include flow in a reaction-diffusion system. We are particularly motivated by the domain of molecular communication, which is the study of communications systems that use molecules as information carriers. While some papers in this area consider mesoscopic models, including \cite{Chou2015,Nakano2010,Gul2010,Wei2013a}, the majority of stochastic simulations (including the simulators presented in \cite{Felicetti2013,Llatser2014,Yilmaz2014a}) use a microscopic approach and are thus unable to take advantage of the scalability of mesoscopic systems. With this in mind, we implement our proposed mesoscopic flow algorithm in the AcCoRD simulator, which is a hybrid microscopic-mesoscopic reaction-diffusion simulator that is publicly available (see \cite{Noel2017a,Noel2016}). However, we emphasize that our proposed method could be applied to \emph{any} implementation of mesoscopic reaction-diffusion that is based on the spatial SSA.

The main contributions of this work are as follows:
\begin{enumerate}
	\item We derive the transition rates for mesoscopic advection-diffusion by solving Fick's laws for one subvolume with steady flow. We perform the derivation along a single dimension (1D) but the results also extend to systems that are 2D and 3D.
	\item We present how to incorporate the advection-diffusion transition rates in the spatial SSA. In particular, we constrain the rates when the flow is sufficiently fast to guarantee that all rates remain non-negative. This integration is implemented in version 1.1 of AcCoRD; see \cite{Noel2016}. The SSA, which implements trajectories as a sequence of events, cannot accommodate a negative transition rate because that would lead to non-causal ``next'' reaction times.
	\item We consider the accuracy of the proposed mesoscopic method when molecules are placed in one subvolume and then observed in another subvolume. We compare the method with a simpler (intuitive) implementation, a microscopic method, and analytical results. We demonstrate that the proposed method is generally more accurate than the simple implementation, and comparable to a microscopic model when the flow is sufficiently slow or the subvolumes are sufficiently small. Both the average and statistical time-varying behavior are considered.
\end{enumerate}

The rest of this paper is organized as follows. We define the system model and present analytical preliminaries in Section~\ref{sec_model}. In Section~\ref{sec_sim_model}, we describe the existing methods for microscopic and mesoscopic simulations. In Section~\ref{sec_meso_flow}, we describe a simple mesoscopic flow method and present our proposed method. All methods are extensively simulated in Section~\ref{sec_results} and we conclude the paper in Section~\ref{sec_end}. 

\section{System and Analytical Models}
\label{sec_model}

In this section, we define the advection-diffusion system of interest, describe Fick's laws of diffusion in the presence of flow, and present analytical results that will be used to assess the accuracy of each simulation method. We do not need to consider chemical reactions because their inclusion does not impact the implementation of advection-diffusion.

\subsection{Physical Environment}

\begin{figure}[!t]
	\centering
	\def\svgwidth{\linewidth}
	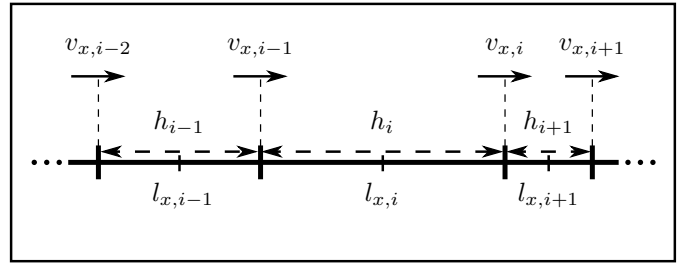
	\caption{Discretized 1-dimensional flow problem. In the most general case, the subvolume length $\subLength{i}$ and flow speed $\flowScalar{\x,i}$ can vary for every subvolume and interface between two adjacent subvolumes, respectively. The coordinate of the center of subvolume $i$, $\posScalar{\x,i}$, is measured relative to the origin.}
	\label{fig_model}
\end{figure}

We consider a single non-reactive chemical species. The environment is partitioned into subvolumes, where the $i$th subvolume has length $\subLength{i}$, is centered at location $\posVec{i} = \posScalar{\x,i}\vec{x} + \posScalar{\y,i}\vec{y} + \posScalar{\z,i}\vec{z}$, and contains $\numMolSubX{i}$ molecules. The subvolumes are placed so that they are flush with Cartesian coordinates, e.g., along the x-dimension in the 1D case in Fig.~\ref{fig_model}, where the center of the $i$th subvolume is $\posScalar{\x,i}$ from the origin.

We permit diffusion and flow to vary over space but not over time (i.e., they are time-invariant but non-uniform). The diffusion coefficient is uniform throughout an individual subvolume and in the $i$th subvolume is $\Dx{i}$. The flow $\flowVec = [\flowScalar{\x,i},\flowScalar{\y,i},\flowScalar{\z,i}]$ is defined at the interface between adjacent subvolumes. In the 1D case, as shown in Fig.~\ref{fig_model}, $\flowScalar{\x,i}$ is the magnitude of the flow from the $i$th subvolume toward the $(i+1)$th subvolume.

\subsection{Fick's Laws of Diffusion}

Fick's laws are equations that are used to derive analytical behavior of diffusion and are also used to formulate the transition rates in mesoscopic simulations. Advection can be directly added to these laws. Fick's first equation in 1D with drift along the $\x$-direction is
\cite[Eq.~(4.4)]{Berg1993}
\begin{equation}
\flux{\x} = -\Dx{}\frac{\partial \Cx{}}{\partial \x} + \flowScalar{x}\Cx{},
\label{eqn_fick_first}
\end{equation}
where $\flux{\x}$ is the flux of molecules along the $\x$-axis, $\Dx{}$ is the diffusion coefficient, $\Cx{}$ is the point molecule concentration, and $\flowScalar{\x}$ is the projection of the flow vector along the $\x$-direction. Eq.~(\ref{eqn_fick_first}) demonstrates that the net flux of molecules at a location $\x$ increases by $\flowScalar{x}\Cx{}$ due to the flow. Fick's second equation in 1D is \cite[Eq.~(4.5)]{Berg1993}
\begin{equation}
\frac{\partial \Cx{}}{\partial t} = \Dx{}\frac{\partial^2 \Cx{}}{\partial \x^2} - \flowScalar{x}\frac{\partial \Cx{}}{\partial \x} = -\boldsymbol{\nabla}\cdot\flux{\x},
\label{eqn_fick_second}
\end{equation}
which demonstrates that, when compared with the no-flow case, the change in concentration at a location is reduced by $\flowScalar{x}\frac{\partial \Cx{}}{\partial \x}$.

\subsection{Analytical Modeling}

In the remainder of this section, we make assumptions that enable the derivation of closed-form expressions for the time-varying concentration or number of molecules at some location. Namely, we assume uniform diffusion $\Dx{}$, uniform flow $\flowVec$, and that the environment is unbounded. Then, if $\Nx{}$ molecules are released at time $\timeX{}=0$ from the origin of the environment, then it can be shown that the \emph{expected} point concentration $\Cxtavg{\dist,t}{}$ along the x-axis at a distance $\dist$ from the origin is
\begin{equation}
\Cxtavg{\dist,t}{} = \frac{\Nx{}}{\left(4\pi \Dx{}
		\timeX{}\right)^\frac{\numDim}{2}}\EXP{-\frac{\left(\dist-\flowScalar{\x}\timeX{}\right)^2 + \flowScalar{\y}^2\timeX{}^2 + \flowScalar{\z}^2\timeX{}^2}{4\Dx{}\timeX{}}},
\label{eqn_passive_response_point}
\end{equation}
where $\numDim$ is the number of dimensions.

In the 1D case, we can integrate (\ref{eqn_passive_response_point}) twice to derive the number of molecules expected over a nonzero-length observer (i.e., a \emph{receiver}) after they are initialized over a nonzero-length source (i.e., a \emph{transmitter}). This result is in closed-form and is convenient for evaluating mesoscopic methods since a mesoscopic environment is described using nonzero-length subvolumes. For clarity of presentation, we assume that the environment is a line of subvolumes with identical length $\subLength{}$, and that the transmitter and receiver  each occupies exactly one subvolume. The transmitter's subvolume is centered around the origin.

We first account for the size of the receiver. The number of molecules expected at the receiver due to a point source can be derived by integrating (\ref{eqn_passive_response_point}) over the receiver's subvolume. This integration is analogous to that used to derive the impulse response at a 3D Cartesian box receiver in the no-flow case in \cite[Eq.~(22)]{Noel2013b}, and relies on substitution and the definition of the error function, i.e., \cite[Eq.~(8.250.1)]{Gradshteyn2007}
\begin{equation}
\label{eqn_erf}
\ERF{\alpha} = \frac{2}{\sqrt{\pi}}\int\limits_0^\alpha \EXP{-\beta^2} \mathrm{d}\beta.
\end{equation}

The result of the integration is the number of molecules expected at the receiver $\Nxtavg{t}{RX}$, and is expressed as
\begin{equation}
\Nxtavg{t}{RX}|_\oneD = \frac{\Nx{}}{2}\left[\ERF{\frac{\xEff+\subLength{}/2}{2\sqrt{\Dx{}\timeX{}}}}
- \ERF{\frac{\xEff-\subLength{}/2}{2\sqrt{\Dx{}\timeX{}}}}\right],
\label{eqn_passive_response_exact_1D}
\end{equation}
where $\xEff=\posScalar{\RX}-\flowScalar{\x}\timeX{}$ and $\posScalar{\RX}$ is the distance from the origin to the center of the receiver.

\addtocounter{equation}{1}
Next, we account for the size of the transmitter centered at the origin. We need to integrate (\ref{eqn_passive_response_exact_1D}) over the transmitter's subvolume (and divide by the length of the subvolume so that the final result has units of molecules). This integration is analogous to the first author's derivation in the 1D no-flow case in \cite[Eq.~(15)]{Noel2016a}, such that the number of molecules expected at the receiver can be expressed by (\ref{volumeCIR_passive_1D}) located at the top of the following page, where
\begin{equation}
\zI = \posScalar{\RX}-\flowScalar{\x}\timeX{} - \subLength{}/2, \qquad \zF = \posScalar{\RX}-\flowScalar{\x}\timeX{} + \subLength{}/2,
\end{equation}
and we use the integral \cite[Eq.~(5.41)]{Gradshteyn2007}
\begin{equation}
\int\ERF{a\beta}\mathrm{d}\beta = \beta\ERF{a\beta} + \frac{1}{a\sqrt{\pi}}\EXP{-a^2\beta^2}.
\label{integration_erf}
\end{equation}

Eq.~(\ref{volumeCIR_passive_1D}) is verbose but it is \emph{exact} and is valid for \emph{any} placement of the receiver relative to the transmitter, including the case where they are in the \emph{same} subvolume, i.e., $\posScalar{\RX}=0$. While it is common to make geometric approximations for the channel behavior, such as the uniform concentration assumption at the receiver (defined in \cite{Noel2013b}) or the point transmitter assumption (defined in \cite{Noel2016a}), having an exact expression is preferred to more precisely measure the accuracy of our proposed simulation method and the alternative approaches.


In the 3D case, an exact closed-form result is not available. However, we briefly present a closed-form approximation, where we assume that the transmitter is a point and not a cube. Thus, we only integrate (\ref{eqn_passive_response_point}) to account for the size of the receiver cube, which is centered along the $\x$-axis. As in the 1D case, this integration is analogous to \cite[Eq.~(22)]{Noel2013b}, such that we correct for flow and write
\begin{align}
\Nxtavg{t}{RX}|_\threeD \approx &\, \frac{\Nx{}}{8}\left[\ERF{\frac{\xEff+\subLength{}/2}{2\sqrt{\Dx{}\timeX{}}}}
- \ERF{\frac{\xEff-\subLength{}/2}{2\sqrt{\Dx{}\timeX{}}}}\right] \nonumber \\
& \times \left[\ERF{\frac{\yEff+\subLength{}/2}{2\sqrt{\Dx{}\timeX{}}}}
- \ERF{\frac{\yEff-\subLength{}/2}{2\sqrt{\Dx{}\timeX{}}}}\right] \nonumber \\
& \times \left[\ERF{\frac{\zEff+\subLength{}/2}{2\sqrt{\Dx{}\timeX{}}}}
- \ERF{\frac{\zEff-\subLength{}/2}{2\sqrt{\Dx{}\timeX{}}}}\right],
\label{eqn_passive_response_approx_3D}
\end{align}
where $\yEff=\posScalar{\RX}-\flowScalar{\y}\timeX{}$ and $\zEff=\posScalar{\RX}-\flowScalar{\z}\timeX{}$.

\begin{figure*}
	\normalsize
	\setcounter{mytempeqncnt}{\value{equation}}
	\setcounter{equation}{5}
	\begin{align}
		\Nxtavg{t}{RX}|_\oneD = &\, \frac{\Nx{}}{\subLength{}}\Bigg\{\sqrt{\frac{\Dx{}\timeX{}}{\pi}}
		\Bigg[\EXP{-\frac{(\zF+\subLength{}/2)^2}{4\Dx{}\timeX{}}} - \EXP{-\frac{(\zF-\subLength{}/2)^2}{4\Dx{}\timeX{}}}
		- \EXP{-\frac{(\zI+\subLength{}/2)^2}{4\Dx{}\timeX{}}} \nonumber \\
		& + \EXP{-\frac{(\zI-\subLength{}/2)^2}{4\Dx{}\timeX{}}}\Bigg] + \frac{1}{2}\Bigg[(\zF + \subLength{}/2)\ERF{\frac{\zF + \subLength{}/2}{2\sqrt{\Dx{}\timeX{}}}} \nonumber \\
		& - (\zI + \subLength{}/2)\ERF{\frac{\zI + \subLength{}/2}{2\sqrt{\Dx{}\timeX{}}}}
		- (\zF - \subLength{}/2)\ERF{\frac{\zF - \subLength{}/2}{2\sqrt{\Dx{}\timeX{}}}} 
		+ (\zI - \subLength{}/2)\ERF{\frac{\zI - \subLength{}/2}{2\sqrt{\Dx{}\timeX{}}}}\Bigg]\Bigg\}
		\label{volumeCIR_passive_1D}
	\end{align}
	\setcounter{equation}{\value{mytempeqncnt}}
	\hrulefill
	\vspace*{4pt}
\end{figure*}

Finally, it is helpful to know when the \emph{peak} number of molecules is expected at the receiver, especially since this time can vary considerably with the strength of the flow. For a molecular communication system, the expected peak observation time will help determine when the receiver should be sampling. For tractability, we consider the peak time of the \emph{point-to-point} concentration in (\ref{eqn_passive_response_point}) when the orthogonal flow components are $\flowScalar{\y} = \flowScalar{\z} = 0$. By taking the derivative of (\ref{eqn_passive_response_point}) with respect to time and setting it equal to 0, it can be shown that the expected peak time $\timeX{\peak}$ in 1D is
\begin{equation}
\timeX{\peak} = \frac{-\Dx{} + \sqrt{\Dx{}^2+\flowScalar{\x}^2\posScalar{\RX}^2}}{\flowScalar{\x}^2},
\label{eqn_tpeak_1D}
\end{equation}
and in 3D is
\begin{equation}
\timeX{\peak} = \frac{-3\Dx{} + \sqrt{9\Dx{}^2+\flowScalar{\x}^2\posScalar{\RX}^2}}{\flowScalar{\x}^2}.
\label{eqn_tpeak_3D}
\end{equation}

\section{Existing Simulation Methods}
\label{sec_sim_model}

In this section, we outline the spatial SSA for diffusion in the absence of flow and the microscopic method with flow. Flow is added to the spatial SSA in the following section.

\subsection{Mesoscopic Method}

Mesoscopic reaction-diffusion simulations go by many different names that refer to the same underlying approach. Some names draw attention to the presence of subvolumes by calling it an \emph{on-lattice} or \emph{compartmental} model; see \cite{Drawert2012,Erban2009,Robinson2015}. Other names focus on the underlying mathematical physics by referring to the \emph{reaction-diffusion master equation} (in \cite{Gillespie2014,Elf2004,Hellander2016}) or the \emph{discrete space continuous time Markov chain} (in \cite{Robinson2015}). It is also common to simply refer to the simulations as mesoscopic; see \cite{Bernstein2005,Engblom2009,Hellander2012}. Throughout this work, we refer to the simulations as mesoscopic and the underlying algorithm as the spatial SSA. Here, we briefly describe the simulation of diffusion in the spatial SSA. Further details of our implementation, including the simulation of chemical reactions, can be found in \cite{Noel2017a}.

In the spatial SSA, the precise location of an individual molecule is uncertain, since all molecules are assumed to be uniformly distributed within a given subvolume. The simulation evolves as a sequence of reaction \emph{events}, where each event is a chemical reaction or the transition of a molecule from one subvolume to an adjacent subvolume. In the absence of chemical reactions, the number of possible events within a subvolume is the number of neighboring subvolumes that share a face. Each event is assigned a propensity $\prop{}$, such that the probability that the event occurs within (infinitesimal) time step $\dtInf$ is $\prop{}\dtInf$. The propensity is the product of the number of molecules $\numMolSubX{}$ in the subvolume and the transition rate $\transRate{}$. In the case of a uniform grid of subvolumes of length $\subLength{}$, the propensity for a molecule to diffuse from subvolume $i$ to neighboring subvolume $j$, $\prop{i,j}$, is  \cite[Eq.~(1.6)]{Flegg2014}
\begin{equation}
\label{eq_meso_diff}
\prop{i,j} = \transRate{i,j}\numMolSubX{i},
\end{equation}
where $\transRate{i,j} = \Dx{i}/\subLength{}^2$ is the transition rate from subvolume $i$ to subvolume $j$. If the subvolumes are non-uniform, then the diffusion transition rate is \cite[Eq.~(15)]{Bernstein2005}
\begin{equation}
\label{eq_meso_diff_1d}
\transRate{i,j} = \frac{2\Dx{i}}{\subLength{i}(\subLength{i} + \subLength{j})}
\end{equation}
in 1D, and \cite[Eq.~(3)]{Noel2017a}
\begin{equation}
\label{eq_meso_diff_3d}
\transRate{i,j} = \frac{2\Dx{i}\areaOverlap}{\subLength{i}^3(\subLength{i} + \subLength{j})}
\end{equation}
in 3D, where $\areaOverlap$ is the size of the shared overlap area between subvolumes $i$ and $j$.

Given the propensity $\prop{i,j}$, the time until the occurrence of a transition from subvolume $i$ to subvolume $j$ can be generated as an exponential random variable via \cite[Eq.~(5)]{Bernstein2005}
\begin{equation}
\label{eq_t_dm}
\timeX{} = -\frac{\ln\uniRV{}}{\prop{i,j}},
\end{equation}
where $\uniRV{}$ is a uniform random number between 0 and 1. There are different implementations to efficiently calculate (\ref{eq_t_dm}) for a large number of dependent events. The implementation in AcCoRD uses the Next Subvolume Method proposed in \cite{Elf2004}.

The primary constraint on the accuracy of the spatial SSA is the subvolume size. When a molecule transitions between subvolumes, it moves from an uncertain location in one subvolume to an uncertain location in the other subvolume. However, decreasing the subvolume size also increases the computational complexity because more subvolumes are needed to model the same total volume, so this is a trade-off.

\subsection{Microscopic Method}
\label{sec_micro}

The microscopic method is particle-based and tracks each molecule individually. The system evolves according to a global time step $\dtMicro$. The displacement of a single molecule is $\normRV{}\sqrt{2\Dx{\indMolType}\dtMicro} + \flowScalar{}\dtMicro$ along each dimension, where $\normRV{}$ is a normal random variable with mean 0 and variance 1, and $\flowScalar{} \in \{\flowScalar{\x},\flowScalar{\y},\flowScalar{\z}\}$ is the flow component along the corresponding dimension. In AcCoRD, collisions with solid environment boundaries that have no surface reactions are reflected perfectly.

\section{Mesoscopic Flow Implementation}
\label{sec_meso_flow}

In this section, we present transition rates for mesoscopic diffusion that account for a net flow or drift. First, we present intuitive transition rates that were implemented in \cite{Drawert2012} in the absence of diffusion. We directly extend these rates to advection-diffusion but this is ultimately naive. Next, we derive the transition rates directly from a discretization of Fick's laws. We present how to modify the spatial SSA so as to include the derived rates, and summarize how flow is implemented in the AcCoRD simulator.

\subsection{Naive Transition Rates}
\label{sec_naive}

A simple (but ultimately naive) approach to modify the mesoscopic transition rate is to only correct the rate in the same direction as the flow. In the absence of flow, the net average displacement of a molecule with unbounded diffusion in a uniform grid of subvolumes is 0, because from (\ref{eq_meso_diff}) the molecule is as likely to diffuse in one direction as in the opposite direction. In the presence of flow, the net average displacement becomes $|\flowVec|/\subLength{}$, where $|\flowVec|$ is the magnitude of the flow. In a 1D environment defined along the $\x$-direction, the simple transition rates for mesoscopic advection-diffusion are
\begin{align}
\label{eqn_rate_against_naive}
\transRate{\mathrm{a}} = &\, \frac{\Dx{}}{\subLength{}^2} - \min\left(0,\frac{\flowScalar{x}}{\subLength{}}\right), \\
\label{eqn_rate_with_naive}
\transRate{\mathrm{w}} = &\, \frac{\Dx{}}{\subLength{}^2} + \max\left(0,\frac{\flowScalar{x}}{\subLength{}}\right),
\end{align}
where $\transRate{\mathrm{a}}$ is the transition rate in the \emph{direction opposite} the flow and $\transRate{\mathrm{w}}$ is the transition rate in the \emph{same direction} as the flow. These simple rates maintain the expected \emph{net} motion but are ultimately ``naive'' because the rate of transition events in the direction against the net flow is the same as in the no-flow case. We will soon demonstrate that this is inaccurate. However, we note that, in the \emph{absence of diffusion}, i.e., $\Dx{}=0$, the simple rates in (\ref{eqn_rate_against_naive}) and (\ref{eqn_rate_with_naive}) as implemented in \cite{Drawert2012} are correct.

\subsection{Derivation of the Mesoscopic Transition Rates}
\label{sec_derivation}

Now we derive the mesoscopic transition rates for advection-diffusion from Fick's laws. Our approach is similar to the derivation of the diffusion rate between non-uniform subvolumes along a one dimensional grid presented in \cite{Bernstein2005}. However, unlike \cite{Bernstein2005}, which considered diffusion alone, we use Fick's laws for diffusion with drift, i.e., (\ref{eqn_fick_first}) and (\ref{eqn_fick_second}). The derivation is presented in 1D, but by orthogonality the results readily apply in 2D and 3D (when subvolumes are in a \emph{uniform} grid; we extend to a more general 3D case in Section~\ref{sec_implementation}).

We begin by writing the 1D differential equation for the number of molecules $\numMolSubX{i}$ in subvolume $i$ \cite[Eq.~(9)]{Bernstein2005}:
\begin{equation}
\frac{\partial \numMolSubX{i}}{\partial t} = -(\transRate{i,i+1} + \transRate{i,i-1})\numMolSubX{i} + \transRate{i+1,i}\numMolSubX{i+1} + \transRate{i-1,i}\numMolSubX{i-1}.
\label{eqn_1d_difference}
\end{equation}

Fick's laws are defined over continuous space and can be used to determine suitable expressions for each $\transRate{}$. To arrive at solutions for the $\transRate{}$s in (\ref{eqn_1d_difference}), we integrate (\ref{eqn_fick_second}) over the $i$th subvolume (centered around the point $\posScalar{\x,i}$; see Fig.~\ref{fig_model}) and apply a linear discretization to the local concentration, i.e., $\Cx{}(\posScalar{\x,i}) = \numMolSubX{i}/\subLength{i}$. Thus, the integration of (\ref{eqn_fick_second}) leads to
\begin{equation}
\frac{\partial \numMolSubX{i}}{\partial t} = \flux{\x}\left(\posScalar{\x,i} - \frac{\subLength{i}}{2}\right) - \flux{\x}\left(\posScalar{\x,i} + \frac{\subLength{i}}{2}\right).
\label{eqn_flux_difference}
\end{equation}

To solve (\ref{eqn_flux_difference}), we require expressions for the flux at each end of the $i$th subvolume. A linear discretization of the concentration at each end is expressed as
\begin{equation}
\Cx{} \left(\posScalar{\x,i} \pm \frac{\subLength{i}}{2}\right) =
\frac{\Cx{} \left(\posScalar{\x,i}\right) + \Cx{} \left(\posScalar{\x,i\pm 1}\right)}{2} =
\frac{1}{2}\left(\frac{\numMolSubX{i}}{\subLength{i}} + \frac{\numMolSubX{i \pm 1}}{\subLength{i \pm 1}}\right).
\label{eqn_end_concentration}
\end{equation}

By discretizing (\ref{eqn_fick_first}) and substituting in (\ref{eqn_end_concentration}), the flux at each end of the $i$th subvolume is
\begin{align}
\flux{\x}\left(\posScalar{\x,i} - \frac{\subLength{i}}{2}\right) = &
-\frac{\Dx{i-1}}{\posScalar{\x,i} - \posScalar{\x,i- 1}}\left(\frac{\numMolSubX{i}}{\subLength{i}} - \frac{\numMolSubX{i-1}}{\subLength{i-1}}\right) \nonumber \\
& + \frac{\flowScalar{x,i-1}}{2}\left(\frac{\numMolSubX{i}}{\subLength{i}} + \frac{\numMolSubX{i-1}}{\subLength{i-1}}\right),
\label{eqn_flux_left}
\end{align}
\begin{align}
\flux{\x}\left(\posScalar{\x,i} + \frac{\subLength{i}}{2}\right) = &
-\frac{\Dx{i}}{\posScalar{\x+1,i} - \posScalar{\x,i}}\left(\frac{\numMolSubX{i+1}}{\subLength{i+1}} - \frac{\numMolSubX{i}}{\subLength{i}}\right) \nonumber \\
& + \frac{\flowScalar{x,i}}{2}\left(\frac{\numMolSubX{i+1}}{\subLength{i+1}} + \frac{\numMolSubX{i}}{\subLength{i}}\right),
\label{eqn_flux_right}
\end{align}
where we recall that $\Dx{i}$ is the coefficient of diffusion of the $i$th subvolume and $\flowScalar{x,i}$ is the flow in the positive $\x$-direction at the interface of the $i$th and $(i+1)$th subvolumes. We can substitute (\ref{eqn_flux_left}) and (\ref{eqn_flux_right}) into (\ref{eqn_flux_difference}) and write
\begin{align}
\frac{\partial \numMolSubX{i}}{\partial t} = &\, \frac{\numMolSubX{i-1}}{\subLength{i-1}} \left(\frac{\Dx{i-1}}{\posScalar{\x,i} - \posScalar{\x,i- 1}} + \frac{\flowScalar{x,i-1}}{2}\right) \nonumber \\
& + \frac{\numMolSubX{i+1}}{\subLength{i+1}} \left(\frac{\Dx{i}}{\posScalar{\x,i+1} - \posScalar{\x,i}} - \frac{\flowScalar{x,i}}{2}\right) \nonumber \\
& - \frac{\numMolSubX{i}}{\subLength{i}} \left(\frac{\Dx{i-1}}{\posScalar{\x,i} - \posScalar{\x,i- 1}} + \frac{\Dx{i}}{\posScalar{\x,i+1} - \posScalar{\x,i}}\right).
\label{eqn_difference_full}
\end{align}

We can readily compare (\ref{eqn_difference_full}) with the differential equation in (\ref{eqn_1d_difference}) to infer the molecule transition rates as
\begin{align}
\label{eqn_rate_right_in}
\transRate{i+1,i} = &\, \frac{1}{\subLength{i+1}} \left(\frac{\Dx{i}}{\posScalar{\x,i+1} - \posScalar{\x,i}} - \frac{\flowScalar{x,i}}{2}\right), \\
\transRate{i-1,i} = &\, \frac{1}{\subLength{i-1}} \left(\frac{\Dx{i-1}}{\posScalar{\x,i} - \posScalar{\x,i- 1}} + \frac{\flowScalar{x,i-1}}{2}\right), \\
\transRate{i,i+1} = &\, \frac{1}{\subLength{i}} \left(\frac{\Dx{i}}{\posScalar{\x,i+1} - \posScalar{\x,i}} + \frac{\flowScalar{x,i}}{2}\right), \\
\label{eqn_rate_left_out}
\transRate{i,i-1} = &\, \frac{1}{\subLength{i}} \left(\frac{\Dx{i-1}}{\posScalar{\x,i} - \posScalar{\x,i- 1}} - \frac{\flowScalar{x,i-1}}{2}\right).
\end{align}

For clarity of presentation, we assume in the remainder of our analysis that the environment is \emph{uniform}, i.e., the subvolumes form a uniform grid of length $\subLength{}$ and are in a fluid with uniform diffusion $\Dx{}$ and flow $\flowScalar{x}$. Under these conditions, the rates $\transRate{\mathrm{a}}$ and $\transRate{\mathrm{w}}$ are against and with the direction of positive flow, respectively, and from (\ref{eqn_rate_right_in})--(\ref{eqn_rate_left_out}) are
\begin{align}
\label{eqn_rate_against}
\transRate{\mathrm{a}} = \frac{\Dx{}}{\subLength{}^2} - \frac{\flowScalar{x}}{2\subLength{}}, \\
\label{eqn_rate_with}
\transRate{\mathrm{w}} = \frac{\Dx{}}{\subLength{}^2} + \frac{\flowScalar{x}}{2\subLength{}}.
\end{align}

Notably, we see that the relation
\begin{equation}
\transRate{\mathrm{w}} - \transRate{\mathrm{a}} = \frac{\flowScalar{x}}{\subLength{}},
\label{eqn_rate_net}
\end{equation}
describes the net motion of molecules and also applies to the naive rates in (\ref{eqn_rate_against_naive}) and (\ref{eqn_rate_with_naive}). The difference in (\ref{eqn_rate_against}) and (\ref{eqn_rate_with}) is that the modified transition rates achieve the net motion by both increasing the no-flow transition rate in the direction of flow and \emph{reducing} the transition rate against the direction of flow.

One might ask whether our task is complete; can we simply replace the transition rates in the no-flow spatial SSA with those in (\ref{eqn_rate_right_in})--(\ref{eqn_rate_left_out})? Unfortunately, we cannot. If the flow component is sufficiently strong, then transition rates against the flow can be negative. Negative rates cannot be accommodated in the SSA, since from (\ref{eq_t_dm}) negative propensities lead to non-causal event times. In the following, we present a mesoscopic method that can accommodate any flow strength.

\subsection{Mesoscopic Method with Flow}
\label{sec_meso_flow_proposed}

Let us consider how to modify the spatial SSA to account for flow. We present an algorithm that sets the transition rates based on the relative strength of the local flow. If the flow and diffusion are \emph{time-invariant}, and if the partitioning of the environment into subvolumes (defined as the set $\setSub$) is unchanged, then we can run the algorithm \emph{a priori} to the simulation (even if $\flowVec$, $\Dx{}$, or $\subLength{}$ are non-uniform). The current number of molecules inside a subvolume does not matter, since the rate describes the likelihood of the behavior of any given molecule. The final algorithm is shown in ``Algorithm~\ref{alg_ssa_rates}.''

\begin{algorithm}[!tb]
	\caption{Determine Spatial SSA Transition Rates}
	\label{alg_ssa_rates}
	\begin{algorithmic}[1]
		\Procedure{Find Transition Rates}{$\Dx{}$, $\subLength{}$, $\flowVec$, $\setSub$} 
		\For{Every pair of adjacent subvolumes in $\setSub$} \label{step_sim_loop}
		\State Identify relevant $\flowScalar{} \in \{\flowScalar{\x},\flowScalar{\y},\flowScalar{\z}\}$
		\State Identify $\transRate{\mathrm{w}}$ and $\transRate{\mathrm{a}}$ according to positive $\flowScalar{}$
		\If{$\flowScalar{} \ge \frac{2\Dx{}}{\subLength{}}$} 
		\State $\transRate{\mathrm{w}} = \flowScalar{}/\subLength{}$ \Comment{Flow is strong and positive}
		\State $\transRate{\mathrm{a}} = 0$
		\ElsIf{$\flowScalar{} \le -\frac{2\Dx{}}{\subLength{}}$}
		\State $\transRate{\mathrm{w}} = 0$
		\State $\transRate{\mathrm{a}} = -\flowScalar{}/\subLength{}$ \Comment{Flow is strong and negative}
		\Else \Comment{Flow is not strong}
		\State $\transRate{\mathrm{w}} = \Dx{}/\subLength{}^2 + \flowScalar{}/(2\subLength{})$
		\State $\transRate{\mathrm{a}} = \Dx{}/\subLength{}^2 - \flowScalar{}/(2\subLength{})$
		\EndIf
		\EndFor
		\EndProcedure
	\end{algorithmic}	
\end{algorithm}

Our primary concern is to avoid instances of negative rates. Otherwise, we should be using the rates derived in Section~\ref{sec_derivation}. Negative transition rates do not occur if the local flow is sufficiently slow, i.e., if
\begin{equation}
|\flowScalar{}| \le \frac{2\Dx{}}{\subLength{}},
\label{eqn_flow_slow}
\end{equation}
in a uniform environment, where we select $\flowScalar{} \in \{\flowScalar{\x},\flowScalar{\y},\flowScalar{\z}\}$ as appropriate for the current interface. We note that the P\'{e}clet number $\Peclet$ for a single subvolume, which defines the strength of advection relative to diffusion, is written as \cite[Eq.~(1.3.1)]{Truskey2009}
\begin{equation}
\Peclet = \frac{|\flowScalar{}|\subLength{}}{\Dx{}},
\label{eqn_peclet}
\end{equation}
so (\ref{eqn_flow_slow}) is analogous to saying that the P\'{e}clet number is less than 2. If (\ref{eqn_flow_slow}) is true, then we use (\ref{eqn_rate_against}) and (\ref{eqn_rate_with}) for $\transRate{\mathrm{a}}$ and $\transRate{\mathrm{w}}$, respectively (or, more generally, the corresponding equations for non-uniform environments).

\begin{figure}[!t]
	\centering
	\def\svgwidth{\linewidth}
	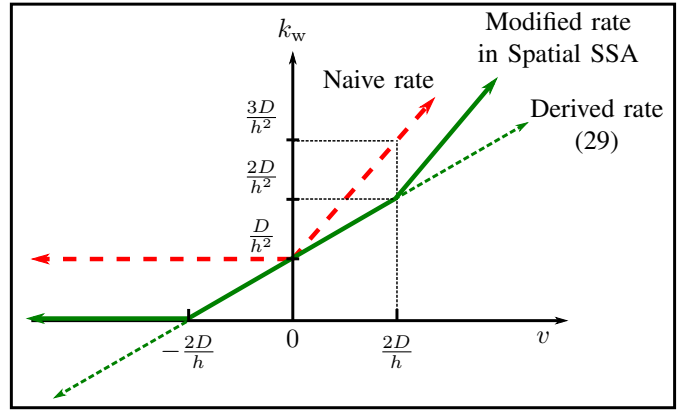
	\caption{Transition rate $\transRate{\mathrm{w}}$ (i.e., in the direction of positive flow) versus flow strength $\flowScalar{}$. The derived rate in (\ref{eqn_rate_with}) has a constant slope of $1/(2\subLength{})$. However, the transition rate in the SSA cannot be negative, so we make it a piecewise constant function of the flow that maintains a constant net flow of molecules across the interface. When $\flowScalar{} < -2\Dx{}/\subLength{}$, the modified rate is $0$. When $\flowScalar{} > 2\Dx{}/\subLength{}$, the modified rate has a slope of $1/\subLength{}$. The naive rate is $\Dx{}/\subLength{}^2$ when $\flowScalar{}\le0$ and has a slope of $1/\subLength{}$ when $\flowScalar{}>0$.}
	\label{fig_rate_vs_v}
\end{figure}

When the local flow is sufficiently large or ``strong'', i.e., when the P\'{e}clet number is 2 or greater, we should still maintain the correct net motion, i.e., $\flowScalar{\x}/\subLength{}$ in (\ref{eqn_rate_net}). In order to do so, and to maintain continuity of the transition rate as a function of the flow speed, we must hold $\transRate{\mathrm{a}} = 0$ and set $\transRate{\mathrm{w}} = \flowScalar{\x}/\subLength{}$ when $\flowScalar{\x} > 2\Dx{}/\subLength{}$, and vice versa when $\flowScalar{\x} < -2\Dx{}/\subLength{}$. These corrections are compared with the naive rates in Fig.~\ref{fig_rate_vs_v} and are also reflected in Algorithm~\ref{alg_ssa_rates}.

Interestingly, from the modified transition rates, there are \emph{no diffusion transitions} in the strong flow regime. What does this mean, and does it make sense? When the flow is sufficiently strong, it is assumed that molecule transitions only occur in the direction of flow. However, the mesoscopic approach assumes that molecules within a subvolume are uniformly distributed. We expect that (occasionally) there will be molecules close to the ``back'' of the subvolume, and these molecules would have a non-negligible probability of diffusing into a subvolume that is against the flow direction, so a complete absence of diffusion transitions is a source of error. Nevertheless, the true locations of molecules within a subvolume are uncertain. We could decrease this uncertainty by keeping track of the number of molecules within subsets of the subvolume, but doing so is analogous to decreasing the subvolume length $\subLength{}$, which from (\ref{eqn_flow_slow}) increases the threshold to be in the strong flow regime. Thus, for a given $\subLength{}$, inaccuracies in the strong flow regime are a side effect of the mesoscopic model that can be mitigated by decreasing the subvolume size.

\subsection{Implementation in 3D Simulations}
\label{sec_implementation}

We have implemented the microscopic and mesoscopic flow methods in the AcCoRD simulator. Flow was introduced in version 1.1; the latest version and select previous versions are available on Github; see \cite{Noel2016}. In the following, we describe how a user can define the flow parameters for a simulation, and present additional modifications to the transition rates to accommodate for 3D subvolumes that are misaligned or have different sizes.

AcCoRD (Actor-based Communication via Reaction-Diffusion) is a hybrid (microscopic-mesoscopic) reaction-diffusion simulation tool. It runs as a standalone executable and the output can be imported and processed in MATLAB. A simulation environment and the corresponding parameters are defined by the user in a configuration file. An environment is composed of \emph{regions}, each of which is either microscopic or mesoscopic. System input and output is achieved by the placement of \emph{actors} within the regions. Each actor is either a source of molecules (which can be modulated by binary data) or an observer of molecules, i.e., either a transmitter or receiver. Chemical reactions (zeroth, first, or second order) and diffusion coefficients can be defined either globally or within individual regions. Additional details on the implementation and use of AcCoRD \edit{up to version 1.0}{R1C3} can be found in \cite{Noel2017a}.

We added flow to \edit{version 1.1 of AcCoRD by enabling}{} a user to specify which types of molecules are able to flow and where they flow. A global 3D flow vector can be defined for the entire environment, and this vector can be replaced within any region. Different region flow vectors can be defined for different types of molecules.

Displacement within a given region is straightforward and follows the descriptions for microscopic flow in Section~\ref{sec_micro} and mesoscopic flow in a uniform environment in Section~\ref{sec_meso_flow_proposed} (since all subvolumes within a region are the same size). Transitions between two microscopic regions use the flow vector that applied to the molecule at the start of the current time step. Transitions between microscopic and mesoscopic regions or vice versa (i.e., at a hybrid interface) follow the diffusion-only transition rules described in \cite{Noel2017a} (re-deriving those rules to account for flow is beyond the scope of this work but can be considered in the future). Transitions between two mesoscopic regions are treated as a special case if the two regions have subvolumes of different sizes \emph{or} if the faces of the subvolumes are misaligned. When this occurs, it is insufficient to apply the rates for inhomogeneous subvolumes in (\ref{eqn_rate_right_in})--(\ref{eqn_rate_left_out}), because they only apply in 1D. Adjacent 1D subvolumes always share a ``face.'' However, in 3D, adjacent subvolumes that are of different sizes or that are misaligned will have an \emph{overlap area} $\areaOverlap$ that is not equal to the areas of both subvolume faces. For diffusion alone, the corrected transition is (\ref{eq_meso_diff_3d}). A similar correction can be applied to the transition rate with advection-diffusion, such that the rate is scaled by the relative overlap area $\areaOverlap/\subLength{i}^2$. By recognizing that the distance between the centers of adjacent subvolumes $i$ and $j$ (along the direction of the shared face) is $(\subLength{i} + \subLength{j})/2$, it can then be shown from (\ref{eqn_rate_right_in})--(\ref{eqn_rate_left_out}) that the corrected transition rate from subvolume $i$ to subvolume $j$ is
\begin{equation}
\label{eq_meso_flow_3d}
\transRate{i,j} = \frac{\areaOverlap}{\subLength{i}^2(\subLength{i} + \subLength{j})}\left[\frac{2\Dx{i}}{\subLength{i}} + \flowScalar{}\right],
\end{equation}
where $\flowScalar{}$ is the flow component that corresponds with the direction from subvolume $i$ to subvolume $j$ (and accounting for the sign as necessary). Eq.~(\ref{eq_meso_flow_3d}) can be analogously included in Algorithm~\ref{alg_ssa_rates}.

As of the time of writing, a flow vector in AcCoRD must be uniform within a given region. A locally-varying flow (e.g., laminar flow, which is predominant in small blood vessels; see \cite{Nelson2008}) can be spatially discretized by defining multiple regions. We note that regions can be placed inside of other regions, so there is still considerable flexibility to define flow in a simulation environment.

\section{Simulation Results}
\label{sec_results}

In this section, we present simulations to demonstrate the accuracy of the proposed mesoscopic advection-diffusion algorithm. We compare the output with the analytical results presented in Section~\ref{sec_model}, the microscopic method summarized in Section~\ref{sec_micro}, and the naive mesoscopic method presented in Section~\ref{sec_naive}. All simulations were completed and processed with the AcCoRD simulator and its post-processing tools, which are all available on Github; see \cite{Noel2016}. Most of our results focus on a (3D) environment that is equivalently 1D, so that closed-form analytical equations are available. However, for completeness, we also consider a 3D environment that cannot be simplified to a 1D equivalent.

Throughout this section, we use a diffusion coefficient of $\Dx{}=10^{-10}\,\frac{\meter^2}{\second}$. The transmitter and receiver each have a width of $1\,\mu\meter$, which is on the order of a typical bacterial cell; see \cite[Ch.~1]{Alberts}. Unless otherwise noted, the mesoscopic subvolume length is also $\subLength{}=1\,\mu\meter$, so the subvolume P\'{e}clet number is 2 when $\flowScalar{}=0.2\,\frac{\meter\meter}{\second}$. We maintain a uniform mesoscopic subvolume size for a given simulation for both clarity of presentation and to focus on the impact of the global subvolume size. Verification of the proposed transition rates between subvolumes of different sizes, as given by (\ref{eq_meso_flow_3d}), is not shown but is consistent with the other results presented in this section.

In order to focus on observations around the expected peak observation time, the receiver always samples the number of molecules within its space every $\timeX{\peak}/30$ seconds over an interval of at least $[0,3\timeX{\peak}]$, where $\timeX{\peak}$ is found using (\ref{eqn_tpeak_1D}) and (\ref{eqn_tpeak_3D}) for the 1D and 3D environments, respectively. Thus, each simulation has at least $M=91$ samples, though for clarity of presentation we do not show every sample time in every plot. \edit{Except where noted,}{} all simulations were repeated $10^3$ times, and all microscopic simulations use the sampling period for the global time step $\dtMicro$.

\subsection{Measuring Accuracy}

Generally, we choose system parameters that push the limits of the proposed method's accuracy. This is not intended to build a case against our own contribution, but to clearly demonstrate the conditions under which our proposed method is suitable as an alternative to a microscopic model (or some other approach). The accuracy of the diffusion-only mesoscopic method is limited by the choice of subvolume size, and we will find that this is also the case for advection-diffusion, as we expect. A related observation is that accuracy will be limited when the P\'{e}clet number $\Peclet$ is high, despite the corrections that we make to simulate strong flow in Algorithm~\ref{alg_ssa_rates}.

In addition to direct comparisons between the simulated time-varying observations and those expected from analytical results, we use two metrics to measure accuracy in the 1D case. First, we measure the relative deviation from the expected value at the peak observation time, i.e.,
\begin{equation}
\label{eqn_error_peak}
\frac{\Nxt{\timeX{\peak}}{\RX}-\Nxtavg{\timeX{\peak}}{\RX}}{\Nxtavg{\timeX{\peak}}{\RX}},
\end{equation}
where $\Nxt{\timeX{}}{\RX}$ is the number of molecules observed by the receiver during the simulation at time $\timeX{}$, and $\Nxtavg{\timeX{}}{\RX}$ is calculated from (\ref{volumeCIR_passive_1D}). We then average (\ref{eqn_error_peak}) over all realizations. Second, we measure the mean of the absolute deviations from the expected values over the \emph{entire} interval $[0,3\timeX{\peak}]$, i.e.,
\begin{equation}
\label{eqn_error_sum}
\frac{1}{M}\sum_{\timeX{}\in[0,3\timeX{\peak}]}^{}\left|\Nxt{\timeX{}}{\RX}-\Nxtavg{\timeX{}}{\RX}\right|,
\end{equation}
for each realization, where $M=91$ is the number of samples over $[0,3\timeX{\peak}]$, and then average over all realizations. These two measures produce distinct but insightful results with which to compare the simulation methods. The expected peak time is arguably the most important single instant for a simulation to be accurate, but is insufficient to describe the overall accuracy.

\subsection{1D Environment}

We first simulate an environment that is a square rod with dimensions $500\,\mu\meter \times 1\,\mu\meter \times 1\,\mu\meter$. The transmitter is placed at the middle of the rod, such that the rod can be assumed to be infinite in length for the timescales considered, and it releases $\Nx{}=200$ molecules at time $\timeX{}=0$. By initializing molecules over an entire $1\,\mu\meter^3$ cube, and observing molecules over another $1\,\mu\meter^3$ cube, this environment is effectively 1D. Thus, the time-varying number of molecules expected at the receiver can be calculated exactly using (\ref{volumeCIR_passive_1D}).

\begin{figure}[!t]
	\centering
	\includegraphics[width=\linewidth]{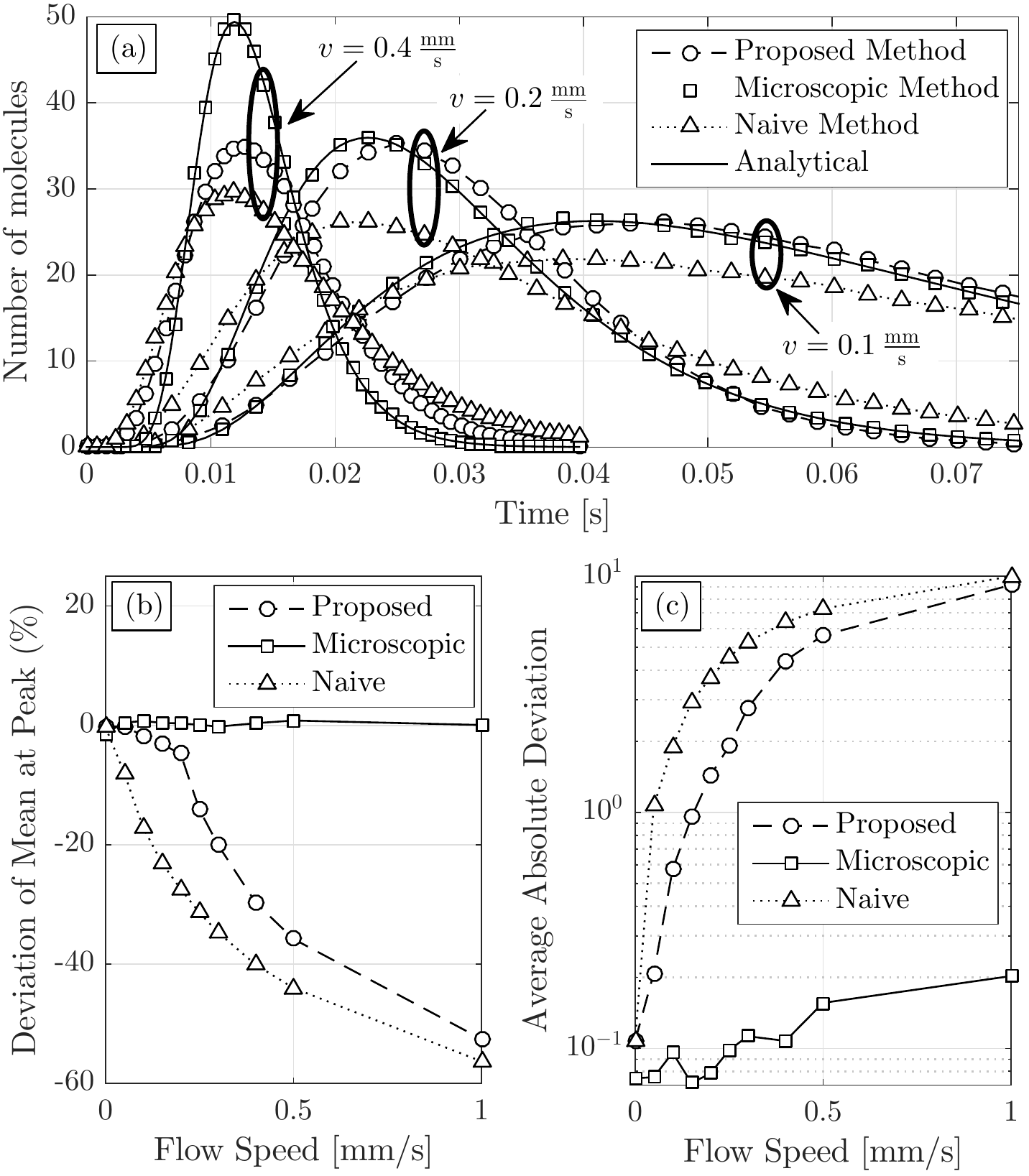}
	\caption{Impact of flow speed $\flowScalar{}$ on the accuracy of simulations when $\posScalar{\RX}=5\,\mu\meter$. (a) The time-varying number of molecules observed for flow speed $\flowScalar{} = \{0.1,0.2,0.4\}\,\frac{\meter\meter}{\second}$, averaged over $10^3$ realizations. (b) Relative deviation of the mean observation at the expected peak time from the expected peak observation, as a function of flow speed. (c) Average absolute deviation of the mean from the expected mean over the range $[0,3\timeX{\peak}]$, as a function of flow speed.}
	\label{fig_meso_avg_vs_time_vary_v}
\end{figure}

In Fig.~\ref{fig_meso_avg_vs_time_vary_v}, we investigate the impact of the flow speed $\flowScalar{}$ on the accuracy of the simulation methods, when the distance separating the centers of the transmitter and receiver is $\posScalar{\RX}=5\,\mu\meter$. In Fig.~\ref{fig_meso_avg_vs_time_vary_v}(a), we measure the average number of molecules observed over time for the flow speeds $\flowScalar{} = \{0.1,0.2,0.4\}\,\frac{\meter\meter}{\second}$. The microscopic simulation matches the analytical curves very well, with no visible deviation. The proposed mesoscopic method shows good agreement when $\flowScalar{} = 0.1\,\frac{\meter\meter}{\second}$ (i.e., $\Peclet=1$), with slight underestimation before the peak time and overestimation after the peak time. Deviations near the peak time are larger at the threshold for strong flow, when $\flowScalar{} = 0.2\,\frac{\meter\meter}{\second}$  (i.e., $\Peclet=2$), and significant for the entire relevant timescale when $\flowScalar{} = 0.4\,\frac{\meter\meter}{\second}$  (i.e., $\Peclet=4$). Nevertheless, the proposed method is visibly more accurate than the naive method, especially when the flow is slow. In Figs.~\ref{fig_meso_avg_vs_time_vary_v}(b) and (c), we measure the error metrics for the peak observation time and the entire observation interval, respectively, as a function of the flow speed. The microscopic method has much less error with high flow speeds than the mesoscopic methods, but we clearly see that the degradation of the proposed method with increasing flow speed is lower than that of the naive method. The deviation of the proposed method at the peak observation time is much closer to the microscopic method than the naive method when $\flowScalar{} \le 0.2\,\frac{\meter\meter}{\second}$, i.e., when $\Peclet\le2$ and we are not in the strong flow regime.

Measuring the time-varying average number of molecules gives an incomplete picture about the fidelity of the simulations, since this does not show the randomness in the stochastic simulations. A deterministic solver should also give the correct \emph{expected} result, \edit{so we also compare the probability \emph{distribution} of the observations with the analytical time-varying probability mass function (PMF). The analytical PMF is based on the fact that, at every observation time, each molecule has the same independent \emph{a priori} probability of being observed by the receiver. So, the total number of molecules observed can be represented as a Binomial random variable $\varOne$ with $\numTrials=\Nx{}$ trials and a success probability $\trialProb$ calculated using (\ref{volumeCIR_passive_1D}) when $\Nx{}=1$. The corresponding PMF, given by \cite[Eq.~(5.1.2)]{Ross2009}
\begin{equation}
\label{eq_pmf_binomial}
\Pr\{\varOne = \numSuccess\} = \binom{\numTrials}{\numSuccess}\trialProb^\numSuccess
\left(1-\trialProb\right)^{\numTrials-\numSuccess},
\end{equation}
can be evaluated at each observation time.}{R2C1}

\begin{figure}[!t]
	\centering
	\includegraphics[width=\linewidth]{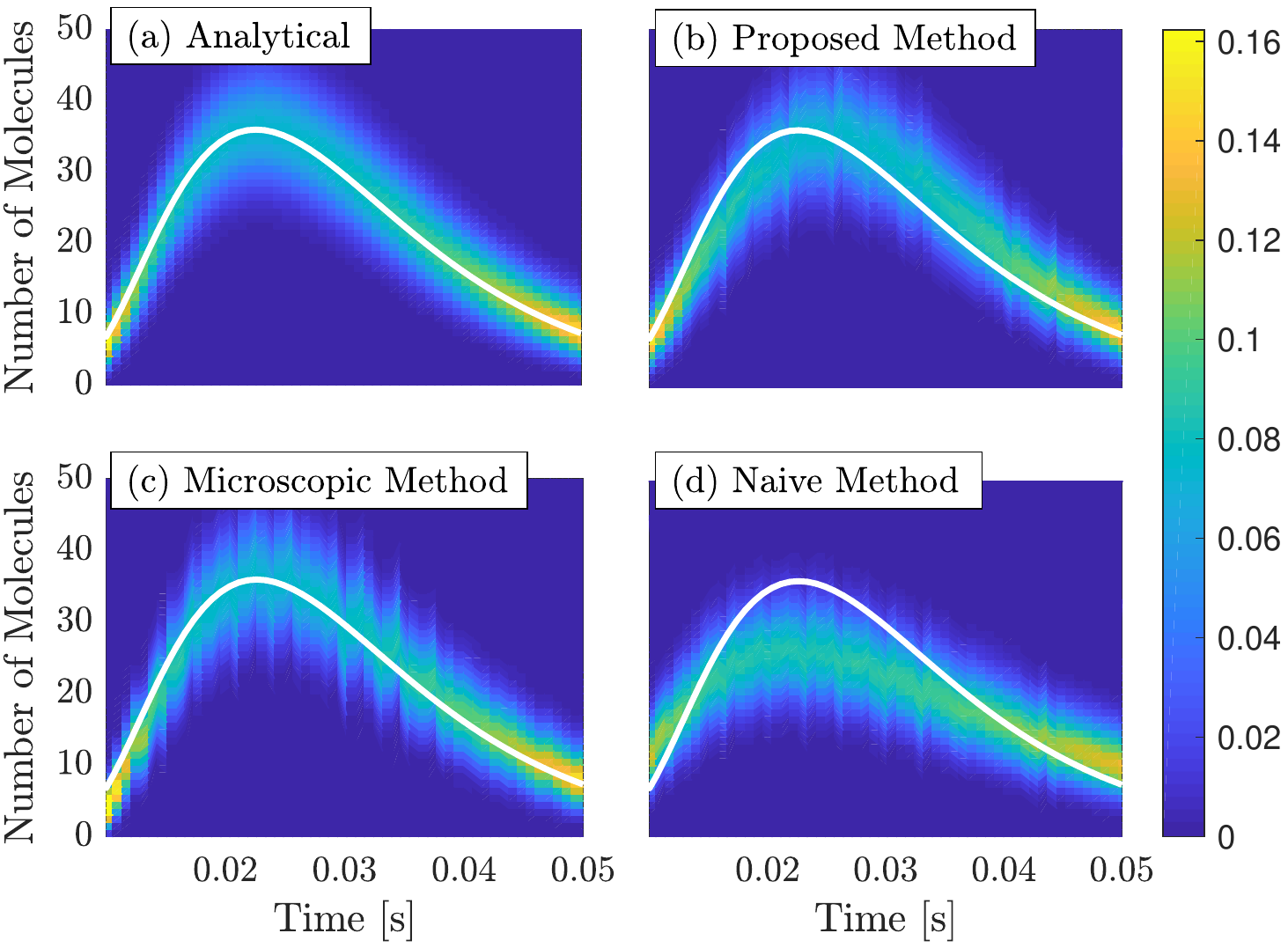}
	\caption{Time-varying PMF from \edit{$5\times10^4$ realizations}{R2C1} of the different simulation methods when $\posScalar{\RX}=5\,\mu\meter$ and $\flowScalar{}=0.2\,\frac{\meter\meter}{\second}$. The domain, range, and distribution scale (shown on the right) of each subplot are the same. The expected analytical response is drawn as a solid white line. (a) The \emph{expected} PMF, calculated as a time-varying Binomial PMF using the expected response (\ref{volumeCIR_passive_1D}). (b) The PMF of the proposed mesoscopic method. (c) The PMF of the microscopic method. (d) The PMF of the naive mesoscopic method.}
	\label{fig_meso_pmf}
\end{figure}

\begin{figure}[!t]
	\centering
	\includegraphics[width=\linewidth]{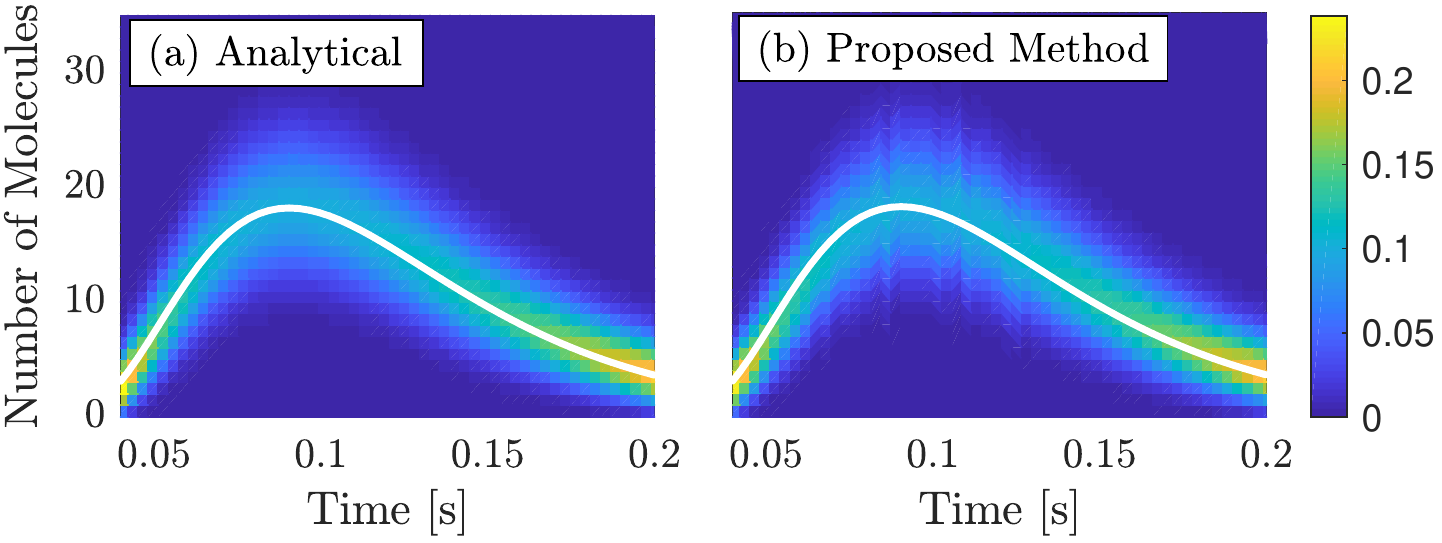}
	\caption{\edit{Time-varying PMF from $5\times10^4$ realizations of the different simulation methods when $\posScalar{\RX}=10\,\mu\meter$ and $\flowScalar{}=0.1\,\frac{\meter\meter}{\second}$. The domain, range, and distribution scale (shown on the right) of each subplot are the same. The expected analytical response is drawn as a solid white line. (a) The \emph{expected} PMF, calculated as a time-varying Binomial PMF using the expected response (\ref{volumeCIR_passive_1D}). (b) The PMF of the proposed mesoscopic method.}{R2C1}}
	\label{fig_meso_pmf_accurate}
\end{figure}

\begin{figure}[!t]
	\centering
	\includegraphics[width=\linewidth]{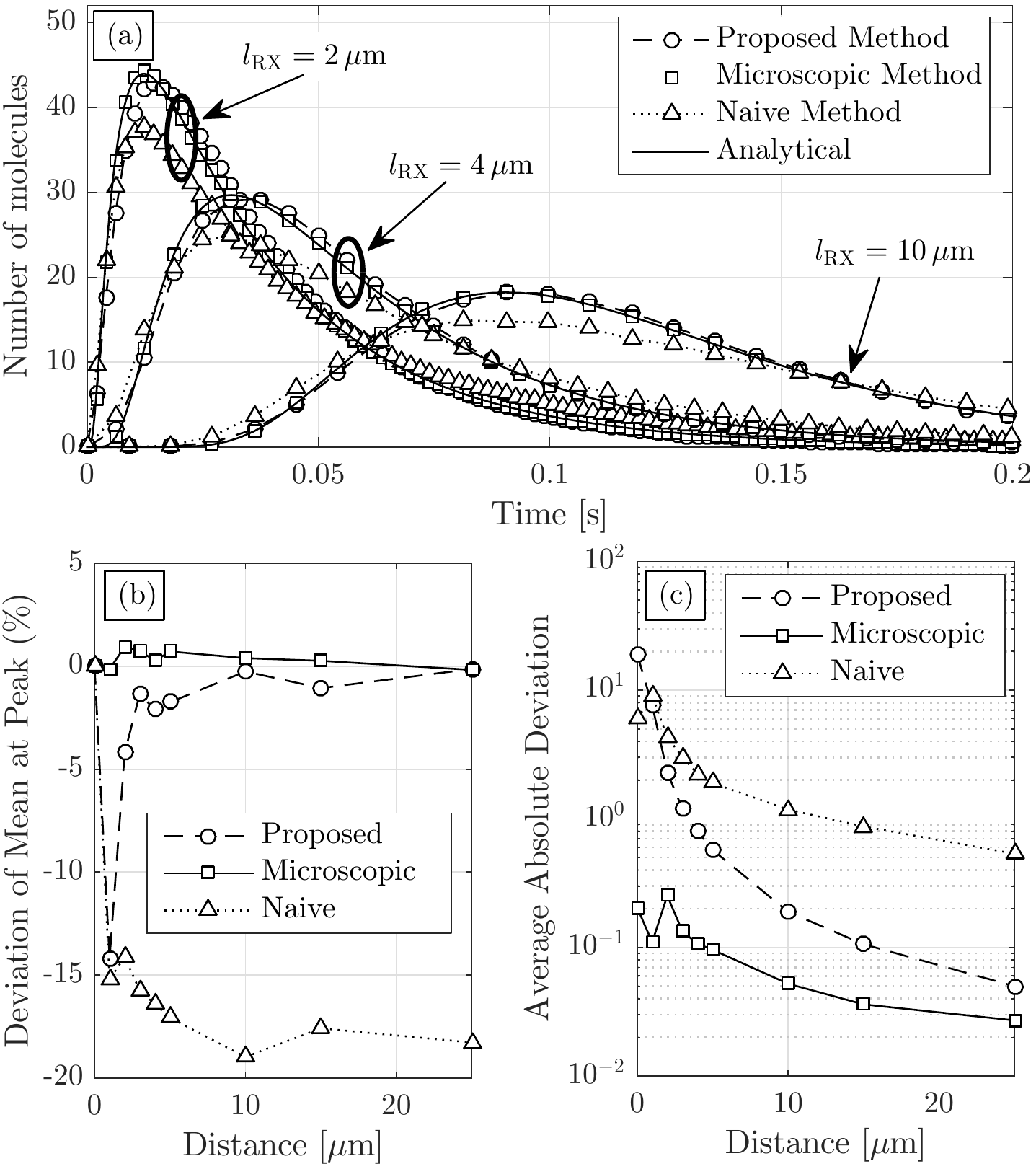}
	\caption{Impact of distance $\posScalar{\RX}$ on the accuracy of simulations when $\flowScalar{}=0.1\,\frac{\meter\meter}{\second}$. (a) The time-varying number of molecules observed at distances of $\posScalar{\RX} = \{2,4,10\}\,\mu\meter$, averaged over $10^3$ realizations. (b) Relative deviation of the mean observation at the expected peak time from the expected peak observation, as a function of distance. (c) Average absolute deviation of the mean from the expected mean over the range $[0,3\timeX{\mathrm{p}}]$, as a function of distance.}
	\label{fig_meso_avg_vs_time_vary_d}
\end{figure}

\edit{We consider the probability distribution of observations at each time step in Figs.~\ref{fig_meso_pmf} and \ref{fig_meso_pmf_accurate}, where we increase the number of simulation realizations to $5\times10^4$}{R2C1}. Specifically, \edit{in Fig.~\ref{fig_meso_pmf} we present the PMF}{R2C1}  for the environment considered for Fig.~\ref{fig_meso_avg_vs_time_vary_v} and where the flow speed is $\flowScalar{} = 0.2\,\frac{\meter\meter}{\second}$. \edit{In Fig.~\ref{fig_meso_pmf_accurate}, we increase the distance to $\posScalar{\RX}=10\,\mu\meter$ and decrease the flow speed to $\flowScalar{}=0.1\,\frac{\meter\meter}{\second}$, which we will see in Fig.~\ref{fig_meso_avg_vs_time_vary_d} leads to better agreement of the proposed approach with the expected analytical behavior. For Fig.~\ref{fig_meso_pmf}, we}{R2C1} \emph{deliberately} chose a flow speed where the proposed approach has deviations from the analytical curve in Fig.~\ref{fig_meso_avg_vs_time_vary_v}, \edit{to demonstrate the statistical integrity of the proposed method even when there are deviations in the expected behavior.}{R2C1}

To facilitate comparisons \edit{within Figs.~\ref{fig_meso_pmf} and \ref{fig_meso_pmf_accurate}}{R2C1}, each subplot includes the \edit{corresponding}{R2C1} expected analytical response drawn as a solid white line, is drawn on the same scales, and has the same distribution values as shown in the color bar legend on the right side of \edit{the figure}{R2C1}. The PMFs shown all begin \edit{with a delay}{R2C1} so that the very high likelihood of observing no molecules at the start of a simulation does not saturate the color bar. Each vertical slice in every subplot is the PMF associated with the corresponding observation time, such that all values associated with a slice add up to 1. \edit{To be considered to have good agreement with the corresponding analytical PMF, a simulation PMF should have the same shape and color. However, every simulation PMF has some ``noisiness'' due to the use of a finite number of realizations.}{R2C1}

Fig.~\ref{fig_meso_pmf} shows the analytical time-varying PMF in Fig.~\ref{fig_meso_pmf}(a). The remaining time-varying PMFs, as shown in Figs.~\ref{fig_meso_pmf}(b), (c), and (d) for the proposed mesoscopic method, microscopic method, and naive mesoscopic method, respectively, are generated from the realizations executed for each method. Despite the visible noise, we can observe the general trends of the PMFs. The microscopic method in \edit{Fig.~\ref{fig_meso_pmf}(c)}{R2C1}, whose average behavior matched the analytical curve very well, also has a time-varying PMF that is very consistent with the analytical time-varying PMF. The deviations in the proposed approach from the analytical curve that we observed in Fig.~\ref{fig_meso_avg_vs_time_vary_v} are also evident here \edit{in Fig.~\ref{fig_meso_pmf}(b)}{R2C1}, with underestimation before the expected peak and overestimation after the expected peak. However, we observe that the general ``shape'' of the statistical distributions with the proposed approach are consistent with the microscopic and analytical distributions. Even the naive method has similar statistical behavior, although the deviations from the other approaches are much more evident.

\edit{In Fig.~\ref{fig_meso_pmf_accurate}, we only compare the analytical time-varying PMF with the PMF corresponding to the proposed method. There is very good agreement here, which is consistent with the improved accuracy under the given physical parameters as we will see in Fig.~\ref{fig_meso_avg_vs_time_vary_d}. Overall, based on Figs.~\ref{fig_meso_pmf} and \ref{fig_meso_pmf_accurate}}{R2C1}, we claim that using a mesoscopic method does not fundamentally alter the simulation statistics. In the remainder of this section, we continue to focus on the deviations in the \emph{expected} behavior.

In Fig.~\ref{fig_meso_avg_vs_time_vary_d}, we consider the impact of the distance $\posScalar{\RX}$ between the centers of the transmitter and receiver when the flow speed is $\flowScalar{}=0.1\,\frac{\meter\meter}{\second}$, i.e., $\Peclet=1$ and the flow is not ``strong.'' Analogously to Fig.~\ref{fig_meso_avg_vs_time_vary_v}, in Fig.~\ref{fig_meso_avg_vs_time_vary_d}(a) we measure the average number of molecules observed over time at the distances $\posScalar{\RX} = \{2,4,10\}\,\mu\meter$. The proposed mesoscopic method agrees very well with both the microscopic method and the analytical curve when the distance is high, i.e., $\posScalar{\RX} = 10\,\mu\meter$. There are slight deviations near the peak when the distance is decreased to $\posScalar{\RX} = 4\,\mu\meter$, and slightly larger deviations at $\posScalar{\RX} = 2\,\mu\meter$. The naive method visibly deviates from the expected behavior at all three distances.

Fig.~\ref{fig_meso_avg_vs_time_vary_d}(b) shows that the proposed method only deviates significantly from the expected peak observation when the distance is very small, i.e., over $14\,\%$ when $\posScalar{\RX} = 1\,\mu\meter$. The naive method deviates from the expected peak observation by at least this amount for all distances $\posScalar{\RX} > 2\,\mu\meter$. We note that there is no deviation in any method when $\posScalar{\RX} = 0$ because the peak time in this case is when the first sample is taken at the start of the simulation (i.e., before molecules have had a non-negligible chance to leave the transmitter's subvolume). Fig.~\ref{fig_meso_avg_vs_time_vary_d}(c) shows that the overall absolute deviation of both mesoscopic methods from the analytical behavior is comparable at short distances, but the proposed method improves and approaches the accuracy of the microscopic method as the distance increases. For $\posScalar{\RX} > 3\,\mu\meter$, the average absolute deviation of the proposed method is less than 1 molecule per sample.

\begin{figure}[!t]
	\centering
	\includegraphics[width=\linewidth]{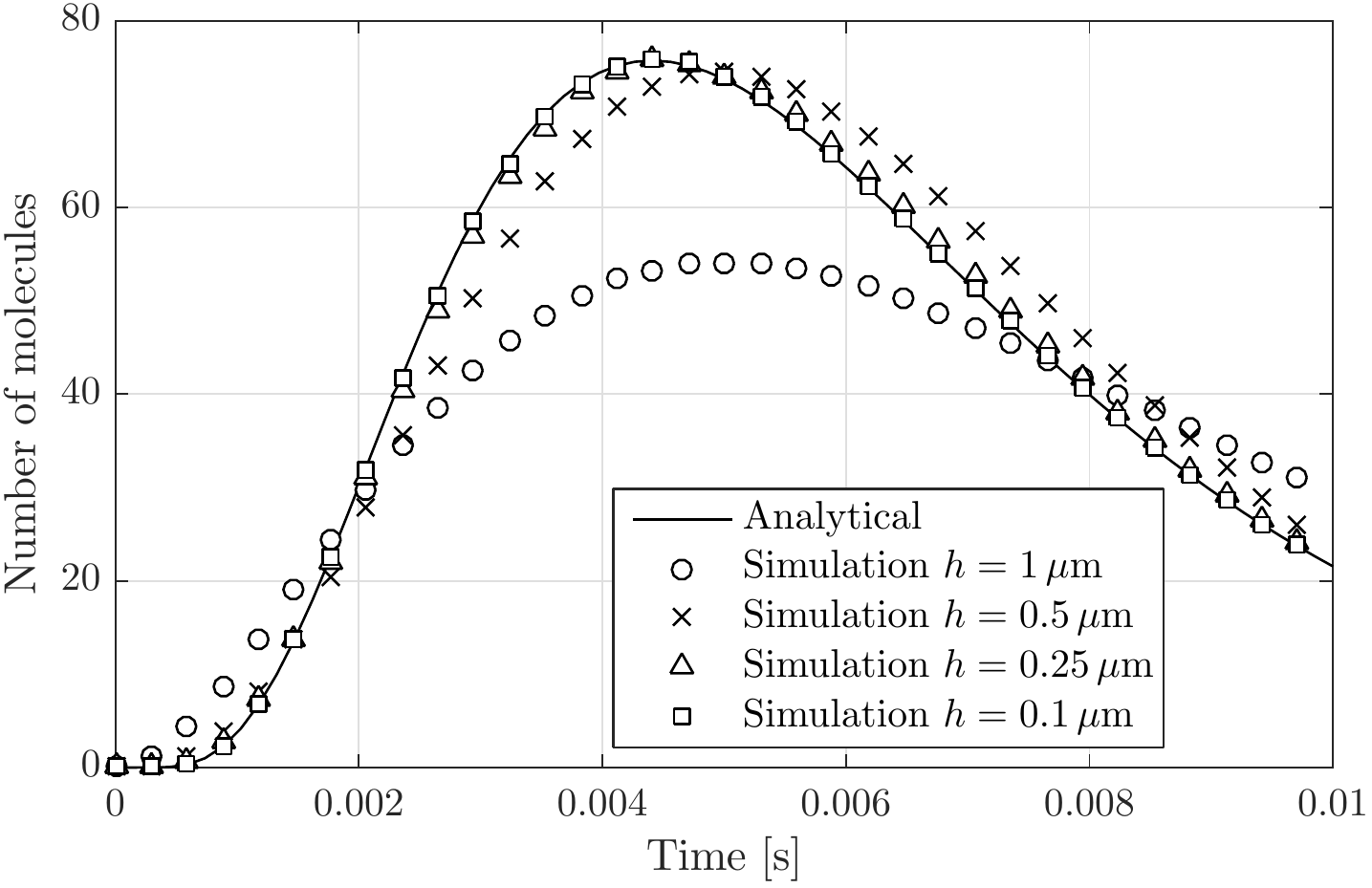}
	\caption{Average number of molecules versus time for the proposed simulation method with varying subvolume size $\subLength{}$ when $\flowScalar{}=0.4\,\frac{\meter\meter}{\second}$ and $\posScalar{\RX}=2\,\mu\meter$. Decreasing the subvolume size decreases the spatial uncertainty that is inherent in a mesoscopic approach.}
	\label{fig_meso_avg_vs_time_vary_h}
\end{figure}

To emphasize that the accuracy of the proposed method is constrained by the subvolume size, we observe the average number of molecules observed over time with the proposed method for different subvolume lengths $\subLength{}$ in Fig.~\ref{fig_meso_avg_vs_time_vary_h}, where the flow speed is $\flowScalar{} = 0.4\,\frac{\meter\meter}{\second}$ and the distance separating the centers of the transmitter and receiver is only $\posScalar{\RX}=2\,\mu\meter$. As we might expect from Figs.~\ref{fig_meso_avg_vs_time_vary_v} and \ref{fig_meso_avg_vs_time_vary_d}, there is significant deviation from the analytical curve with the default subvolume size of $\subLength{}=1\,\mu\meter$, but the accuracy clearly improves with decreasing $\subLength{}$. When $\subLength{}=0.1\,\mu\meter$, such that $\Peclet=0.4$, the average observations with the proposed method are practically indistinguishable from the analytical curve. The trade-off with this value of $\subLength{}$ is that we need 20 times more subvolumes along each linear dimension than in the default case, which increases the simulation runtime by a comparable amount. More details on the computational complexity of the simulation methods, and the implementation in the AcCoRD simulator in particular, can be found in \cite{Noel2017a}.

\subsection{3D Environment}

Finally, we present results to demonstrate the suitability of our proposed mesoscopic method for 3D environments that are not effectively 1D. We simulate an environment that is a cube of width $21\,\mu\meter$. The transmitter is placed at the center of the cube and releases $\Nx{}=10^4$ molecules at time $\timeX{}=0$. The receiver is placed so that its center is $\posScalar{\RX}=5\,\mu\meter$ from the center of the transmitter and so that the faces of the transmitter and receiver cubes are parallel. The linear dimensions of this environment are much smaller than the previous environment, but we still assume that they are large enough for the system to be unbounded. A closed-form exact solution for the number of molecules expected at the receiver is unavailable, but we refer to (\ref{eqn_passive_response_approx_3D}), which is the number of molecules expected due to a \emph{point} source, as an analytical reference. For clarity of presentation, we do not consider the naive mesoscopic model for this environment, but its accuracy is comparable to how it performed for the 1D environment.

\begin{figure}[!t]
	\centering
	\includegraphics[width=\linewidth]{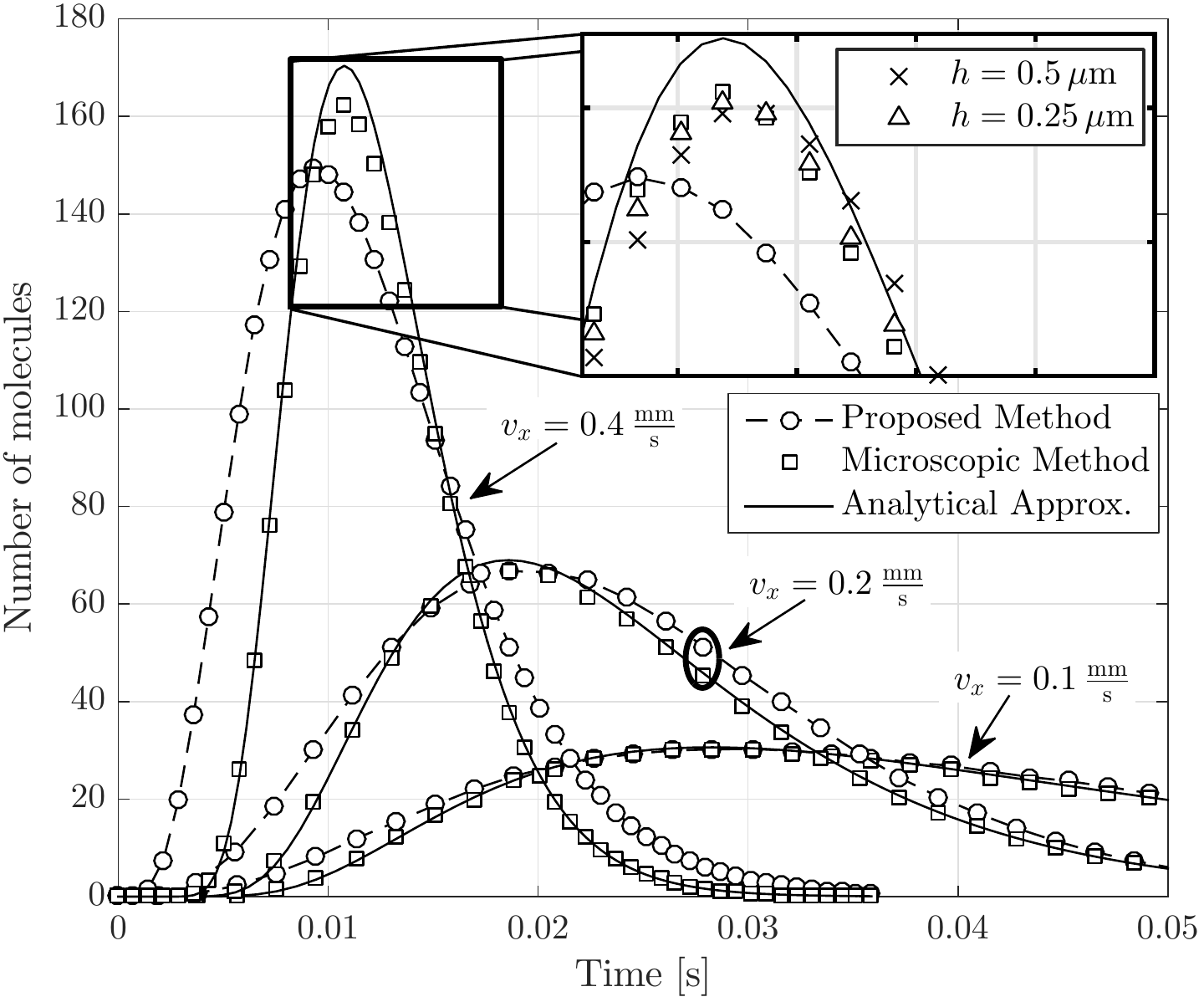}
	\caption{The time-varying number of molecules observed for flow speed $\flowScalar{\x} = \{0.1,0.2,0.4\}\,\frac{\meter\meter}{\second}$, averaged over $10^3$ realizations of the 3D environment. The analytical response is for a point transmitter, whereas the simulations release molecules over the volume of a transparent cube. The inset shows that the channel response of the proposed mesoscopic method can approach that of the microscopic method by decreasing the subvolume size $\subLength{}$.}
	\label{fig_meso_avg_vs_time_3D_vary_v}
\end{figure}

In Fig.~\ref{fig_meso_avg_vs_time_3D_vary_v}, we measure the average number of molecules observed over time for the flow speeds $\flowScalar{\x} = \{0.1,0.2,0.4\}\,\frac{\meter\meter}{\second}$, where the $\x$-direction is along the line from the transmitter to the receiver. We set $\flowScalar{\y} = \flowScalar{\z} =0$. Unlike Fig.~\ref{fig_meso_avg_vs_time_vary_v}, the microscopic simulations with faster flow speeds deviate from the analytical curves, which only serve as approximations. Thus, it is more important to compare the proposed mesoscopic method with the microscopic model. When the flow speed is only $\flowScalar{\x}=0.1\,\frac{\meter\meter}{\second}$, there is good agreement with the microscopic method near the peak, but there are visible deviations at other observation times (particularly when the curves are increasing). Deviations of the proposed method from the microscopic method are observed for most samples when the flow speed is $\flowScalar{\x}=0.2\,\frac{\meter\meter}{\second}$, and are very large for $\flowScalar{\x}=0.4\,\frac{\meter\meter}{\second}$, such that the diffusion wave clearly arrives sooner with the proposed method.

Very large deviations in the proposed method from the microscopic method are not encouraging, but we show in the inset of Fig.~\ref{fig_meso_avg_vs_time_3D_vary_v} that they are due to the constraints on the mesoscopic approach and not the proposed method itself. The inset focuses on the expected peak time when $\flowScalar{\x}=0.4\,\frac{\meter\meter}{\second}$ and adds the average observations of the proposed method when the subvolume size is decreased to $\subLength{}=0.5\,\mu\meter$ and $\subLength{}=0.25\,\mu\meter$ (corresponding to $\Peclet=2$ and $\Peclet=1$, respectively). There is a clear improvement in the accuracy as the subvolume size decreases, as we also observed for the 1D environment in Fig.~\ref{fig_meso_avg_vs_time_vary_h}.

\begin{figure}[!t]
	\centering
	\includegraphics[width=\linewidth]{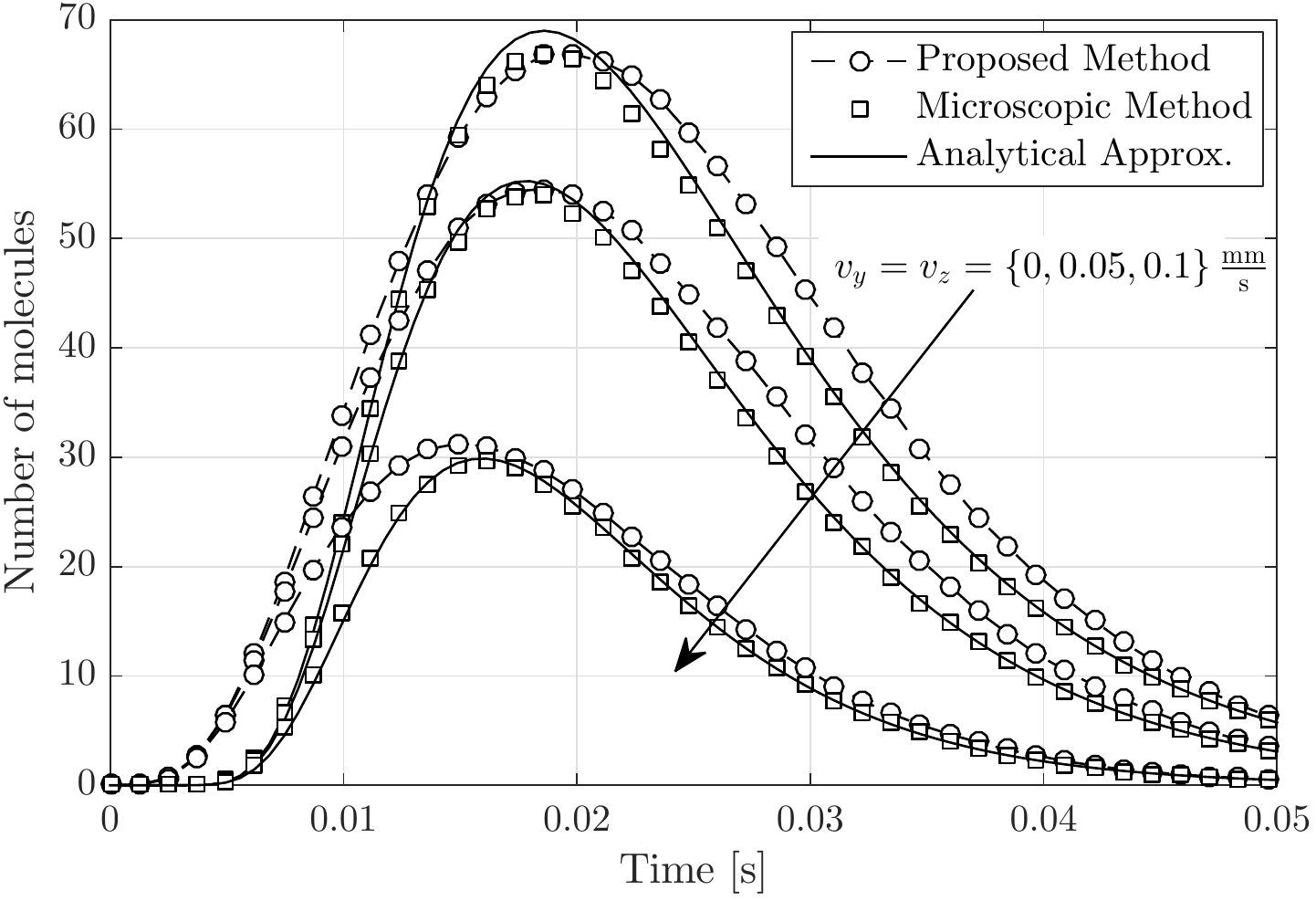}
	\caption{The time-varying number of molecules observed for flow component $\flowScalar{\x} = 0.2\,\frac{\meter\meter}{\second}$, averaged over $10^3$ realizations of the 3D environment. The other flow components are $\flowScalar{\y} = \flowScalar{\z} = \{0,0.05,0.1\}\,\frac{\meter\meter}{\second}$.}
	\label{fig_meso_avg_vs_time_3D_vary_v3D}
\end{figure}

All flows considered thus far have only been in the direction from the transmitter to the receiver and have not had orthogonal components. To demonstrate that our proposed method is also valid with flow along more than one dimension, in Fig.~\ref{fig_meso_avg_vs_time_3D_vary_v3D} we measure the average number of molecules observed over time for the flow components $\flowScalar{\y} = \flowScalar{\z} = \{0,0.05,0.1\}\,\frac{\meter\meter}{\second}$. We keep the flow component $\flowScalar{\x}=0.2\,\frac{\meter\meter}{\second}$ and do not alter the sampling times from the corresponding scenario in Fig.~\ref{fig_meso_avg_vs_time_3D_vary_v}. The orthogonal flow components are sufficient to substantially change the number of molecules observed, but the deviations between the two methods remain comparable. Thus, we are confident that the proposed method is also suitable for flow in any direction.

\section{Conclusions}
\label{sec_end}

In this paper, we derived the transition rates between adjacent mesoscopic subvolumes in the presence of flow and diffusion. We integrated Fick's second law in 1D to observe how the transition rates change in the presence of flow, and extended the results to apply in 3D. The transition rates were implemented in the hybrid microscopic-mesoscopic simulator AcCoRD by ensuring that the rates remain non-negative. Simulation results demonstrated that the derived transition rates are accurate for advection-diffusion systems when the chosen subvolumes are sufficiently small (i.e., with P\'{e}clet number sufficiently less than 2).

Future opportunities to extend the contributions of this work include the integration of the proposed method with other mesoscopic algorithms, including the scalability offered by spatial ``tau leaping''. \edit{Tau-leaping enables a simulation to progress as either purely mesoscopic, purely continuous, or via an intermediate approach such as the Langevin method, depending on the current local state of the system; see \cite{Iyengar2010}}{R2C2}. Furthermore, the proposed method could be more rigorously integrated with hybrid microscopic-mesoscopic transitions. The approach in this work might also be applied to the examination of other forms of biased transport, such as in the presence of external electromagnetic fields.

\bibliography{../../../references/library_fixed}

\end{document}